\documentclass[preprintnumbers,twocolumn,aps,unsortedaddress,showpacs,amsmath]{revtex4}

\usepackage{graphics,graphicx}
\preprint{Physical Review E}

\begin{document}

\title{Vortex tubes in turbulence velocity fields at Reynolds numbers Re$_{\lambda} \simeq 300$--$1300$}

\author{Hideaki Mouri}
\email{hmouri@mri-jma.go.jp}

\author{Akihiro Hori}
\altaffiliation[Also at ]{Meteorological and Environmental Service, Inc., Tama, Tokyo 206-0012, Japan.}

\author{Yoshihide Kawashima}
\altaffiliation[Also at ]{Meteorological and Environmental Service, Inc., Tama, Tokyo 206-0012, Japan.}
\affiliation{Meteorological Research Institute, Nagamine 1-1, Tsukuba 305-0052, Japan}

\date{July 20, 2004}

\begin{abstract}
The most elementary structures of turbulence, i.e., vortex tubes, are studied using velocity data obtained in a laboratory experiment for boundary layers with Reynolds numbers Re$_{\lambda} = 295$--1258. We conduct conditional averaging for enhancements of a small-scale velocity increment and obtain the typical velocity profile for vortex tubes. Their radii are of the order of the Kolmogorov length. Their circulation velocities are of the order of the root-mean-square velocity fluctuation. We also obtain the distribution of the interval between successive enhancements of the velocity increment as the measure of the spatial distribution of vortex tubes. They tend to cluster together below about the integral length and more significantly below about the Taylor microscale. These properties are independent of the Reynolds number and are hence expected to be universal.
\end{abstract}

\pacs{47.27.Ak, 47.27.Jv, 47.27.Nz}

\maketitle

\section{INTRODUCTION}
\label{s1}

Turbulence contains vortex tubes as the most elementary structures \cite{f95,sa97,s99,kida}. Regions of strong vorticity are organized into tubes. The energy dissipation is significant around them. They occupy a small fraction of the volume and are embedded in the background flow that is random and of large scales. Their existence has been established at microscale Reynolds numbers Re$_{\lambda} \alt 2000$, by seeding a turbulent liquid with gas bubbles and thereby visualizing regions of low pressure that are associated with strong vorticity \cite{d91,cdc95,lvmb00,vsg95}.

Direct numerical simulations \cite{vm91,j93,tmi99,kida} have derived basic parameters of vortex tubes at low Reynolds numbers, Re$_{\lambda} \alt 200$. The radii are of the order of the Kolmogorov length $\eta$. The total lengths are of the order of the integral length $L$. The circulation velocities are of the order of the root-mean-square velocity fluctuation $\langle u^2 \rangle ^{1/2}$ or the Kolmogorov velocity $u_K$. Here $\langle \cdot \rangle$ denotes an average. The lifetimes are of the order of the large-eddy turnover time $L/ \langle u^2 \rangle ^{1/2}$.

The universality of these tube parameters has not been established because their behavior at high Reynolds numbers has not been known. It is difficult to conduct a direct numerical simulation at Re$_{\lambda} \agt 200$ \cite{k03}. Bubble visualization does not provide sufficient information \cite{note0}. We accordingly use velocity fields of laboratory turbulence at high Reynolds numbers to study some of the tube parameters. The velocity field is intermittent at small scales. A small-scale velocity variation is enhanced at the positions of vortex tubes \cite{cdc95,b96,n97,cg99,mtk99,mhk00,mt02}.

There are several possible configurations for laboratory experiments. Although the most popular configuration is isotropic grid turbulence, its Reynolds number is not high \cite{mhk00}. We instead use rough-wall boundary-layer turbulence. The highest Reynolds number achieved in our experiment is Re$_{\lambda} = 1258$, which exceeds those in almost all the previous studies for vortex tubes in the velocity field \cite{b96}.

Using a probe suspended in the flow, we obtained a one-dimensional cut of the velocity field. We measured not only the velocity component in the mean-flow direction but also the component that is perpendicular to the mean-flow direction. The latter component is suited to detecting circulation flows such as those associated with vortex tubes \cite{n97,cg99,mtk99,mhk00,mt02}.

The experiment is described in Sec.~\ref{s2}. We present a model for vortex tubes in Sec.~\ref{s3}. From the experimental data, the typical velocity profile for vortex tubes is extracted and its radius and circulation velocity are obtained in Sec.~\ref{s4}. The spatial distribution of vortex tubes is obtained in Sec.~\ref{s5}. The dependences of these tube parameters on the Reynolds number are studied in Sec.~\ref{s6}. We conclude with remarks in Sec.~\ref{s7}.
\begingroup
\squeezetable
\begin{table*}
\caption{\label{t1} 
Summary of experimental conditions, flow characteristics, and parameters of vortex tubes. The kinematic viscosity reflects the air temperature in the wind tunnel. The velocity derivative was obtained as $\partial _x v$ = $[ 8 v(x+ \delta x)- 8 v(x- \delta x)-v(x+ 2 \delta x)+v(x- 2 \delta x)]/ 12 \delta x$ with $\delta x = U/f_s$.}

\begin{ruledtabular}
\begin{tabular}{lllcccccc}
\multicolumn{2}{l}{quantity}                                                                                         & unit                         & \multicolumn{6}{c}{values for individual data sets}\\ 
\hline
 experimental conditions:         &                                                                                  &                              &        &        &        &        &        &        \\  
 incoming-wind velocity           & $U_i$                                                                            & m s$^{-1}$                   & 2      & 4      & 8      & 12     & 16     & 20     \\
 sampling frequency               & $f_s$                                                                            & kHz                          & 4      & 8      & 16     & 24     & 32     & 40     \\
 kinematic viscosity              & $\nu$                                                                            & cm$^2$ s$^{-1}$              & 0.144  & 0.145  & 0.145  & 0.147  & 0.150  & 0.149  \\
 99\% thickness                   &                                                                                  & cm                           & 78     & 80     & 80     & 79     & 79     & 79     \\
 displacement thickness           & $\int_0^{\hat{z}} (1-U/\hat{U}) dz$                                              & cm                           & 20     & 23     & 23     & 23     & 22     & 22     \\
 measurement height               &                                                                                  & cm                           & 35     & 35     & 30     & 30     & 25     & 25     \\
\hline
 flow characteristics:            &                                                                                  &                              &        &        &        &        &        &        \\  
 mean streamwise velocity         & $U$                                                                              & m s$^{-1}$                   & 1.59   & 3.08   & 5.83   & 8.81   & 11.1   & 13.8   \\
 streamwise flatness factor       & $\langle u^4 \rangle / \langle u^2 \rangle ^2$                                   &                              & 2.73   & 2.68   & 2.70   & 2.69   & 2.70   & 2.71   \\
 spanwise flatness factor         & $\langle v^4 \rangle / \langle v^2 \rangle ^2$                                   &                              & 3.03   & 3.03   & 3.00   & 3.06   & 3.01   & 3.02   \\ 
 energy dissipation rate          & $\langle \varepsilon \rangle = 15 \nu \langle (\partial _x v )^2 \rangle /2$     & m$^2$ s$^{-3}$               & 0.0316 & 0.254  & 2.03   & 5.65   & 14.6   & 26.2   \\
 streamwise velocity fluctuation  & $\langle u^2 \rangle ^{1/2}$                                                     & m s$^{-1}$                   & 0.271  & 0.554  & 1.15   & 1.71   & 2.38   & 2.95   \\
 spanwise velocity fluctuation    & $\langle v^2 \rangle ^{1/2}$                                                     & m s$^{-1}$                   & 0.227  & 0.462  & 0.958  & 1.42   & 2.01   & 2.53   \\
 Kolmogorov velocity              & $u_K = ( \nu \langle \varepsilon \rangle )^{1/4}$                                & m s$^{-1}$                   & 0.0260 & 0.0438 & 0.0737 & 0.0955 & 0.122  & 0.141  \\
 streamwise integral length       & $L_u = \int \langle u(x+\delta x) u(x) \rangle / \langle u^2 \rangle d \delta x$ & cm                           & 48.4   & 44.2   & 39.0   & 42.3   & 43.6   & 43.6   \\
 spanwise integral length         & $L_v = \int \langle v(x+\delta x) v(x) \rangle / \langle v^2 \rangle d \delta x$ & cm                           & 7.76   & 7.08   & 7.06   & 6.23   & 5.79   & 5.92   \\
 Taylor microscale                & $\lambda = [ 2 \langle v^2 \rangle / \langle (\partial _x v )^2 \rangle ]^{1/2}$ & cm                           & 1.88   & 1.35   & 0.991  & 0.889  & 0.788  & 0.740  \\
 Kolmogorov length                & $\eta = (\nu ^3 / \langle \varepsilon \rangle )^{1/4}$                           & cm                           & 0.0554 & 0.0331 & 0.0197 & 0.0154 & 0.0123 & 0.0106 \\
 microscale Reynolds number       & Re$_{\lambda} = \langle v^2 \rangle ^{1/2} \lambda / \nu$                        &                              & 295    & 430    & 655    & 861    & 1054   & 1258   \\
\hline
 parameters of vortex tubes:      &                                                                                  &                              &        &        &        &        &        &        \\  
 radius                           & $R_0$                                                                            & $\eta$                       & 6.08   & 6.04   & 6.28   & 7.14   & 6.98   & 7.37   \\
 circulation velocity             & $V_0$                                                                            & $\langle v^2 \rangle ^{1/2}$ & 0.600  & 0.526  & 0.485  & 0.476  & 0.469  & 0.464  \\
 circulation velocity             & $V_0$                                                                            & $u_K$                        & 5.23   & 5.55   & 6.31   & 7.10   & 7.73   & 8.37   \\
 Reynolds number                  & Re$_0 = V_0 R_0 / \nu$                                                           &                              & 32     & 34     & 40     & 51     & 54     & 62     \\
 clustering scale                 & $\delta x_0$                                                                     & $\lambda$                    & 2.38   & 2.90   & 2.82   & 2.77   & 2.67   & 2.92   \\
 probability density              & $P_0(\lambda)$                                                                   & $\lambda^{-1}$               & 0.349  & 0.451  & 0.536  & 0.508  & 0.523  & 0.545  \\
\end{tabular}
\end{ruledtabular}
\end{table*}
\endgroup

\section{EXPERIMENT}
\label{s2}

The experiment was done in a wind tunnel of the Meteorological Research Institute. We use the coordinates $x$, $y$, and $z$ in the streamwise, spanwise, and floor-normal directions. The corresponding wind velocities are $u$, $v$, and $w$. The origin $x = y = z = 0$ is taken on the tunnel floor at the entrance to the test section. Its size was $\delta x = 18$ m, $\delta y = 3$ m, and $\delta z = 2$ m. Over the entire floor of the test section, we placed blocks as roughness elements. Their size was $\delta x = 6$ cm, $\delta y = 21$ cm, and $\delta z = 11$ cm. The spacing of adjacent blocks was $\delta x = \delta y = 0.5$ m. We set the incoming-wind velocity to be $U_i = 2$, 4, 8, 12, 16, or 20 m~s$^{-1}$.

The streamwise and spanwise velocities were simultaneously measured using a hot-wire anemometer. The anemometer was composed of a crossed-wire probe and a constant temperature system. The wires were made of platinum-coated tungsten, 5 $\mu$m in diameter, 1.25 mm in effective length, 1 mm in separation, 280\,$^{\circ}$C in temperature, and oriented at $\pm 45 ^{\circ}$ to the streamwise direction. The calibration was done before and after each of the measurements. We did not measure the floor-normal velocity, which suffers from mean shear and hence is less suited to studying vortex tubes than the spanwise velocity.

The measurement positions were at $x = 12.5$ m, where the boundary layer was well developed. The 99\% thickness, i.e., the height at which the mean streamwise velocity $U$ is 99\% of its maximum value $\hat{U}$, was 0.8 m. The displacement thickness $\textstyle \int_0^{\hat{z}} (1-U/\hat{U}) dz$ was 0.2 m. Here $\hat{z}$ is the height for the velocity $\hat{U}$ \cite{note1}. The 99\% thickness and displacement thickness were independent of the incoming-wind velocity. Thus, among all the measurements, the overall flow structure was the same. The reason is that the Reynolds number for the entire boundary layer was high enough.

We determined the measurement height $z$ so that the flatness factor for the spanwise velocity $\langle v^4 \rangle / \langle v^2 \rangle ^2$ was close to the Gaussian value of 3.  The flatness factor was less than 3 at small heights where the flow was affected by the surface roughness. The flatness factor was greater than 3 at large heights where the flow was affected by the fluctuation of the interface to the outer laminar flow. We obtained the Gaussian value of 3 at an intermediate height, where eddies with various sizes and strengths filled the space randomly and independently \cite{mthk03}. There the flatness factor for the streamwise velocity $\langle u^4 \rangle / \langle u^2 \rangle ^2$ was different from 3 because the turbulence was not isotropic at large scales.

The signal was low-pass filtered at $f_c$ = 2--20 kHz with 24 dB per octave and sampled digitally at $f_s$ = 4--40 kHz with 16-bit resolution. To avoid aliasing, the sampling frequency was set to be twice the filter cutoff frequency, $f_s = 2 f_c$. The data length was $2 \times 10^7$ points for the incoming-wind velocities $U_i = 4$, 8, and 12 m s$^{-1}$. It was $8 \times 10^7$ points for the incoming-wind velocities $U_i = 2$, 16, and 20 m s$^{-1}$. 

\begin{figure}[!]
\resizebox{8.4cm}{!}{\includegraphics*[3cm,9cm][17cm,22cm]{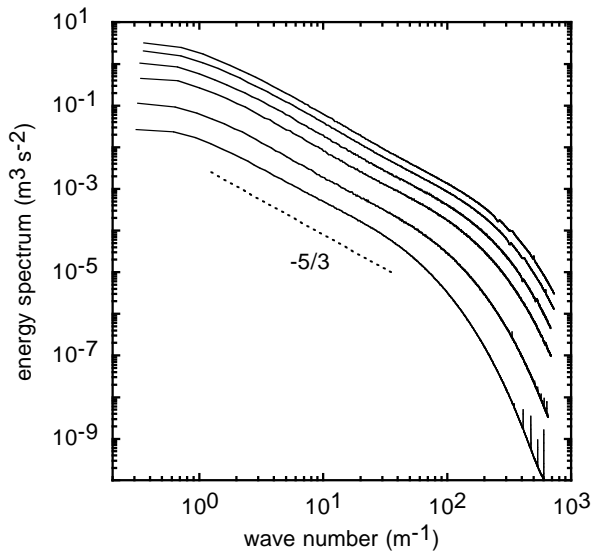}}
\caption{\label{f1} Energy spectrum of the spanwise velocity at Re$_{\lambda} = 295$, 430, 655, 861, 1054, and 1258 (from bottom to top). The wave number $k$ is in units of m$^{-1}$ instead of the usual radian m$^{-1}$. The dotted line denotes Kolmogorov's $k^{-5/3}$ law for the inertial range.}
\end{figure}

The energy spectra of the spanwise velocity are shown in Fig. \ref{f1}. We have used Taylor's frozen-eddy hypothesis to convert temporal variations into spatial variations in the streamwise direction. Throughout the energy spectra, the signal-to-noise ratio is high.

The experimental conditions and flow characteristics are summarized in Table \ref{t1}. The microscale Reynolds number Re$_{\lambda}$ ranges from 295 to 1258. We have obtained the smallest-scale statistics from the spanwise-velocity gradient $\partial_x v$ instead of the usual streamwise-velocity gradient $\partial_x u $, by assuming the smallest-scale isotropy $\langle ( \partial_x v )^2 \rangle = 2 \langle ( \partial_x u )^2 \rangle$. This is because, especially at small scales in strong turbulence, the $u$ component measured by a crossed-wire probe is contaminated with the $w$ component that is perpendicular to the plane of the two wires of the probe \cite{b95,note2}. The $v$ component is free from such contamination. For the smallest-scale isotropy, we have no direct evidence. The smallest-scale isotropy still serves as a meaningful assumption, even if it was not achieved, because our present results are then comparable with those obtained in isotropic turbulence.

\section{MODEL FOR VORTEX TUBES}
\label{s3}

The representative model for vortex tubes is the Burgers vortex, an axisymmetric steady circulation in a strain field. In cylindrical coordinates, the circulation and the strain field are written, respectively, as
\begin{subequations}
\begin{eqnarray}
\label{eq1a}
& & 
  u_{\Theta} \propto \frac{\nu}{a R} 
             \left[ 1 - \exp \left( - \frac{a R^2}{4 \nu} \right) \right]
  \qquad (a > 0), \\
\label{eq1b}
& & 
\left( u_R , u_{\Theta}, u_Z \right) =
\left( - \frac{1}{2} a R, 0, a Z \right).
\end{eqnarray}
\end{subequations}
Here $\nu$ is the kinematic viscosity. The circulation is maximal at $R$ = $R_0$ = $2.24 (\nu / a)^{1/2}$. Thus $R_0$ is regarded as the tube radius.

Suppose that the axis of the vortex tube penetrates the $(x,y)$ plane at the point $(0,\Delta)$. The $x$ and $y$ axes are, respectively, in the streamwise and spanwise directions. If the direction of the tube axis is $(\theta, \varphi)$ in spherical coordinates, the streamwise ($u$) and spanwise ($v$) components of the circulation flow $u_{\Theta}$ along the $x$ axis are
\begin{subequations}
\begin{eqnarray}
\label{eq2a}
&u(x)& = \frac{\Delta \cos \theta}{R} u_{\Theta} (R), \\
\label{eq2b}
&v(x)& = \frac{x \cos \theta}{R} u_{\Theta} (R),
\end{eqnarray}
\end{subequations}
with
\begin{eqnarray}
\label{eq3}
R^2 &=& x^2  ( 1 - \sin ^2 \theta \cos ^2 \varphi )        
        +\Delta ^2 ( 1 - \sin ^2 \theta \sin ^2 \varphi )  \nonumber \\
    & & +2x \Delta \sin ^2 \theta \sin \varphi \cos \varphi.
\end{eqnarray}
For the radial inflow $u_R$ of the strain field, the streamwise and spanwise components are
{\footnotesize
\begin{subequations}
\begin{eqnarray}
\label{eq4a}
&u(x)& = \frac{x ( 1 - \sin ^2 \theta \cos ^2 \varphi ) 
               + \Delta \sin ^2 \theta \sin \varphi \cos \varphi }
              {R} 
         u_R (R), \\
\label{eq4b}
&v(x)& = - \frac{x \sin ^2 \theta \sin \varphi \cos \varphi 
                 + \Delta ( 1 - \sin ^2 \theta \sin ^2 \varphi )}
                {R} 
           u_R (R). \quad
\end{eqnarray}
\end{subequations}
}If the vortex tube passes close to the probe ($\Delta \alt R_0$) and the tube is not heavily inclined ($\theta \simeq 0$), the spanwise velocity is dominated by the small-scale circulation flow [Eq. (\ref{eq2b})]. The streamwise velocity is dominated by the large-scale radial inflow [Eq. (\ref{eq4a})]. This situation is of our interest. The velocity profiles of vortex tubes with $\Delta \alt R_0$ and $\theta \simeq 0$ are nearly the same \cite{b96}. If $\Delta \gg R_0$ or $\theta \gg 0$, the tube signal is weak at least in the spanwise velocity at small scales.

Since there are large-scale velocity fluctuations, a vortex tube in actual turbulence does not necessarily pass the probe along the mean streamwise direction \cite{b96}. This fact was nevertheless not serious in our experiment. The incident angle $\phi$ of the vortex tube was not large, i.e., $\langle \phi \rangle \simeq \arctan ( \langle v ^2 \rangle ^{1/2} / U ) \lesssim 10 ^{\circ}$ for $\langle v ^2 \rangle ^{1/2} / U \lesssim 0.2$ as in Table \ref{t1}.

\section{VELOCITY PROFILE OF VORTEX TUBES}
\label{s4}

\begin{figure}[!]
\resizebox{8.4cm}{!}{\includegraphics*[3.5cm,4cm][18cm,26cm]{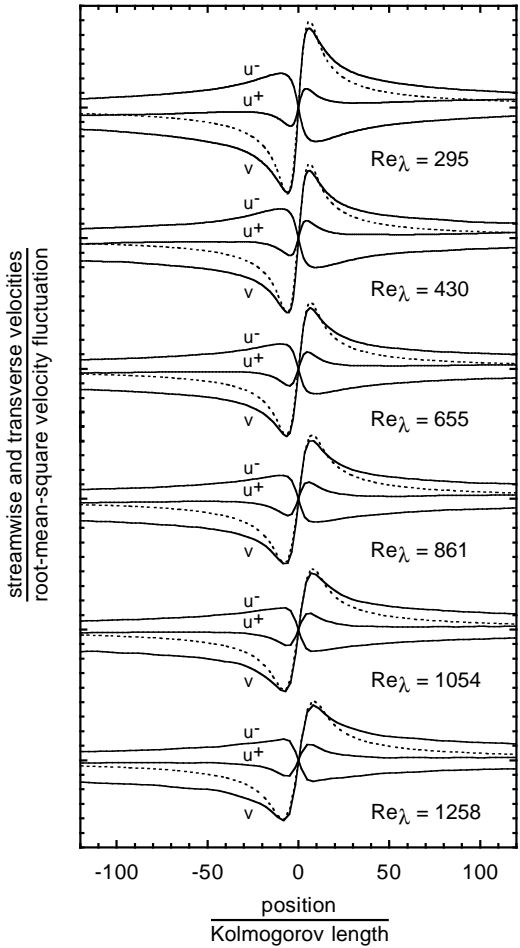}}
\caption{\label{f2} Typical profiles for vortex tubes in the streamwise ($u$) and spanwise ($v$) velocities at Re$_{\lambda} = 295$, 430, 655, 861, 1054, and 1258 (from top to bottom). The $u$ profile is shown separately for $u(x+ U/f_s) - u(x) > 0$ and $u(x+ U/f_s) - u(x) \le 0$ at $x = 0$ (designated as $u^+$ and $u^-$). The position $x$ is normalized by the Kolmogorov length $\eta$. On the ordinate, the unit scale corresponds to one-tenth of the root-mean-square velocity fluctuation $\langle v^2 \rangle ^{1/2}$. The $v$ profiles of Burgers vortices with $\Delta = 0$ and $\theta = 0$ are shown with dotted lines [see Eqs. (\ref{eq2b}) and (\ref{eq5})].}
\end{figure}

The typical profiles for vortex tubes in the streamwise ($u$) and spanwise ($v$) velocities are extracted by averaging signals centered at the position where the absolute value of the spanwise-velocity increment $\vert v(x+ \delta x) - v(x) \vert$ is enhanced above a certain threshold \cite{cg99,mtk99,mhk00}. We set the scale $\delta x$ to be the sampling interval $U/f_s$. The threshold is set to be the highest percentile for the absolute values of the velocity increments. Thus 1\% of them are used for the averaging. When the velocity increment is negative, we invert the sign of the $v$ signal before the averaging. The results are shown in Fig. \ref{f2}.

\begin{figure}[!]
\resizebox{8.4cm}{!}{\includegraphics*[3.5cm,4cm][18cm,26cm]{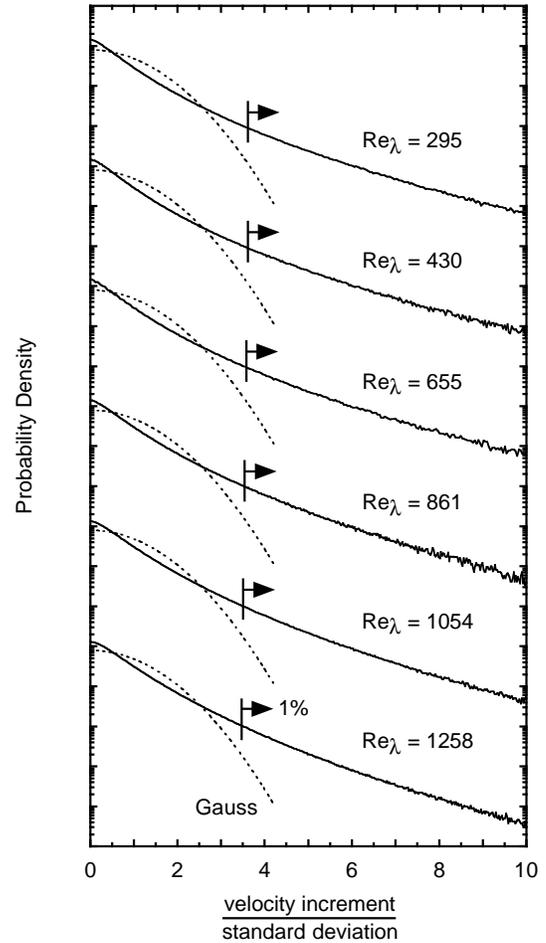}}
\caption{\label{f3} Probability density distribution of the absolute velocity increment $\vert v(x+ U/f_s) - v(x) \vert$ at Re$_{\lambda} = 295$, 430, 655, 861, 1054, and 1258 (from top to bottom). The distribution is shifted vertically by a factor $10^3$. The velocity increment is normalized by its standard deviation, which is 0.00678, 0.0186, 0.0497, 0.0830, 0.124, and 0.167 m s$^{-1}$ (from top to bottom). The arrows indicate the range of the enhanced velocity increments used in our analyses, which share 1\% of the total. The dotted lines denote a Gaussian distribution.}
\end{figure}

The threshold value for the enhancement of the velocity increment has been determined with a compromise. If the threshold is higher, the statistical uncertainty is more significant. If the threshold is lower, the contamination with the background flow is more significant. We nevertheless expect that our following results are qualitatively independent of the threshold if the fraction of the velocity increments used for the averaging is $\lesssim 1$\%. They comprise the tail of the probability density distribution that is well above the Gaussian distribution with the same standard deviation as shown in Fig. \ref{f3} \cite{note3}. The only deficit is that the threshold is too high for some weak vortex tubes. They are not considered here.

The $u$ profiles in Fig. \ref{f2} are separated for $u(x+ \delta x) - u(x) > 0$ and $u(x+ \delta x) - u(x) \le 0$ at $x = 0$ (designated as $u^+$ and $u^-$). We have decomposed the individual profiles into the symmetric and antisymmetric components and have shown only the antisymmetric components. The contamination with the $w$ component leads to a symmetric positive excursion in the $u$ profile \cite{note2,s02}. Such an excursion was seen in our previous study \cite{mhk00,note5}. The $u$ profile averaged for vortex tubes is antisymmetric (see Sec.~\ref{s3}).

To ensure statistical significance, we have reduced the sample size by a factor of 2 and redone the conditional averaging. The velocity profiles do not differ from those in Fig. \ref{f2} by more than a line thickness.

For reference, we show the $v$ profile (\ref{eq2b}) of a Burgers vortex with $\Delta = 0$ and $\theta = 0$ in Fig. \ref{f2} (dotted lines). The radius $R_0$ and the maximum circulation velocity $V_0$ have been determined so as to fit the observed profile around its peaks. There our conditional averaging prefers vortex tubes with $\Delta \simeq 0$ and $\theta \simeq 0$ because such tubes have the strongest signals. Since the Kolmogorov length $\eta$ is smaller than the probe size in the streamwise direction $l$, we have assumed that the measured velocity $v_m(x)$ is the true velocity $v_t(x)$ averaged over the probe size \cite{b95}:
\begin{equation}
\label{eq5}
v_m(x) = \frac{1}{l} \int^{l/2}_{-l/2} v_t(x+x') dx'
\quad
{\rm with}
\quad
l = \frac{1.25\,{\rm mm}}{\sqrt{2}}.
\end{equation} 
The $R_0$ and $V_0$ values are shown in Table \ref{t1}. While the radius $R_0$ is several times the Kolmogorov length $\eta$, the circulation velocity $V_0$ is about a half of the root-mean-square velocity fluctuation $\langle v^2 \rangle ^{1/2}$ or several times the Kolmogorov velocity $u_K$ \cite{vm91,j93,tmi99,kida,b96,n97,mtk99,mhk00,mt02}. 

The observed $v$ profile is close to the profile of a Burgers vortex \cite{tmi99,mhk00,mt02}. We thus confirm the existence of vortex tubes and their responsibility for small-scale intermittency at high microscale Reynolds numbers. The observed $v$ profile has more pronounced tails than the profile of a Burgers vortex. There should be a contribution from vortex tubes that are heavily inclined to the streamwise direction with $\theta \gg 0$ \cite{mhk00}.

The $u^{\pm}$ profiles are dominated by the circulation flow $u_{\Theta}$ of vortex tubes passing the probe at some distances $\Delta > 0$ [Eq. (\ref{eq2a})] or with some incident angles $\phi > 0$ \cite{b96}. If the radial inflow $u_R$ of the strain field were predominant, we would observe $u(x) \simeq -ax/2$ [Eq. (\ref{eq4a})]. This is not the case. The $u^-$ profile only has a somewhat larger amplitude than the $u^+$ profile. Unlike a Burgers vortex, an actual vortex tube is not always oriented in the stretching direction \cite{vm91,j93,mhk00,mt02,t92}.

\section{Spatial distribution of vortex tubes}
\label{s5}

\begin{figure}[!]
\resizebox{8.4cm}{!}{\includegraphics*[3.5cm,4cm][18cm,26cm]{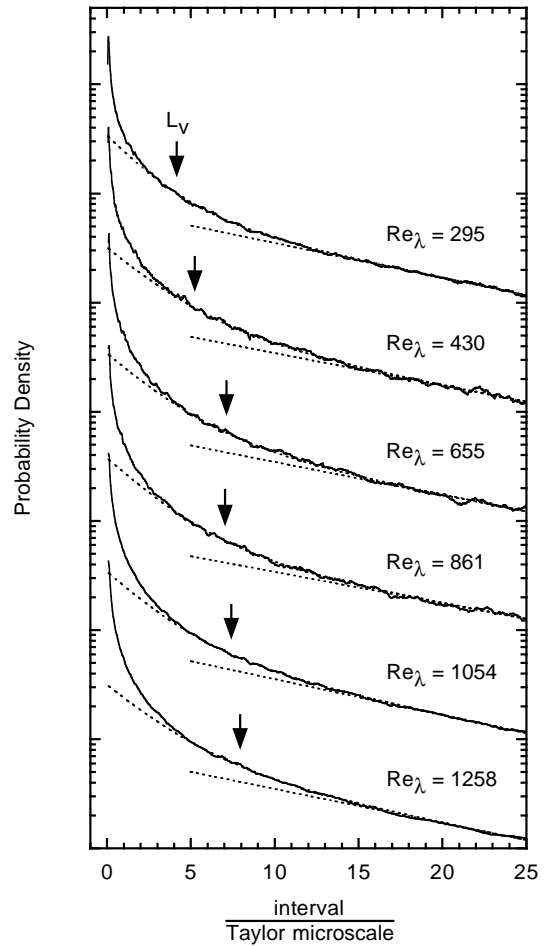}}
\caption{\label{f4} Probability density distribution of the interval $\delta x'$ between vortex tubes at Re$_{\lambda} = 295$, 430, 655, 861, 1054, and 1258 (from top to bottom). The probability density distribution is normalized by the amplitude of its exponential tail (dotted straight lines), and it is shifted vertically by a factor 10. The interval is normalized by the Taylor microscale $\lambda$. The dotted curves denote the results of a least-squares fit with a sum of two exponential functions at $\delta x' = 5\lambda$--$25\lambda$ [Eq. (\ref{eq6})]. The arrows denote the spanwise integral length $L_v$.}
\end{figure}

The spatial distribution of vortex tubes on a one-dimensional cut of a turbulent flow is studied using the probability density $P_0$ of the interval $\delta x'$ between successive enhancements of the spanwise-velocity increment. The enhancement is defined with the same threshold as in Sec.~\ref{s4}. We show the probability density distribution in Fig.~\ref{f4}, where the interval is normalized by the Taylor microscale $\lambda$ in order to cover both the small and large intervals. The statistical significance is less than that of the velocity profiles in Fig.~\ref{f2}, but it is still sufficient for our analysis \cite{note4}.

For intervals $\delta x' \gtrsim 5 \lambda$, we successfully model the probability density with a sum of two exponential functions (dotted curves):
\begin{equation}
\label{eq6}
P_0(\delta x') = c_0 \exp \left( - \frac{\delta x'}{\delta x_0} \right) 
               + c_1 \exp \left( - \frac{\delta x'}{\delta x_1} \right) ,
\end{equation}
with $\delta x_0 < \delta x_1$. The second term implies that the probability density distribution has an exponential tail that appears linear on a semi-logarithmic plot (dotted straight lines). This is characteristic of the Poisson process of random and independent events \cite{f68}. Thus the large-scale distribution of vortex tubes is random and independent. The first term implies that, with decreasing interval, the probability density becomes enhanced over that for the exponential distribution \cite{cdc95,lvmb00,cg99,mhk00}. Since the enhancement is significant below about the spanwise integral length $L_v$, this is attributable to clustering of vortex tubes below the energy-containing scale. It was actually demonstrated in direct numerical simulations that strong vortex tubes lie on the borders of energy-containing eddies \cite{j93}. The clustering scale $\delta x_0$ is a few times the Taylor microscale $\lambda$ (Table \ref{t1}).

With decreasing interval $\delta x'$ below about the scale $\delta x_0$, the probability density becomes enhanced over our model (\ref{eq6}). The clustering of vortex tubes becomes significant. In Table \ref{t1}, we show the probability density at $\delta x' = \lambda$, which has been normalized by the amplitude of the exponential tail. The probability density at the smaller intervals is not  so useful because a very strong vortex tube could cause more than one enhancement of the velocity increment.

\section{DEPENDENCE ON REYNOLDS NUMBER}
\label{s6}

\begin{figure}[!]
\resizebox{8.4cm}{!}{\includegraphics*[3.5cm,5cm][18cm,23cm]{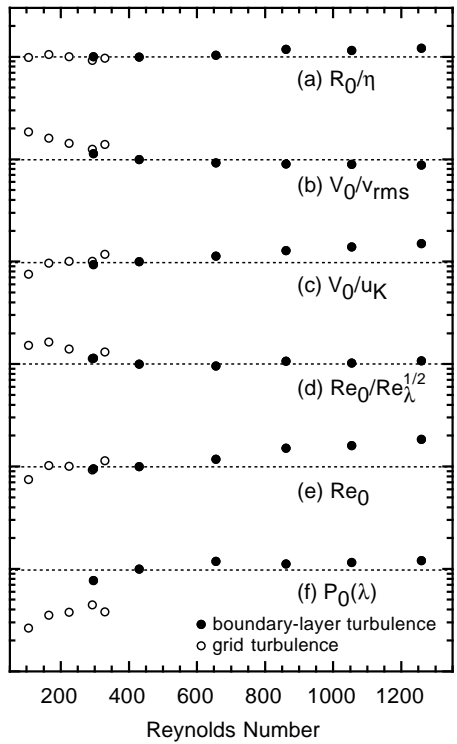}}
\caption{\label{f5} Dependence of tube parameters on Re$_{\lambda}$. (a) $R_0/\eta$. (b) $V_0/\langle v^2 \rangle ^{1/2}$. (c) $V_0/u_K$. (d) Re$_0$/Re$_{\lambda}^{1/2}$. (e) Re$_0$. (f) $P_0(\lambda)$. They are individually normalized by the values at Re$_{\lambda} = 430$. The filled circles denote the data at Re$_{\lambda} = 295$, 430, 655, 861, 1054, and 1258 from our present experiment. The open circles denote the data at Re$_{\lambda} = 105$, 165, 225, 292, and 329 from our previous experiment of grid turbulence \cite{mhk00}. }
\end{figure}

Thus far we have obtained the parameters of vortex tubes, i.e., the radius $R_0$, the maximum circulation velocity $V_0$, and the interval distribution $P_0$. Their dependences on the microscale Reynolds number Re$_{\lambda}$ are studied here. To extend the Re$_{\lambda}$ range, we also use velocity data from our previous experiment of grid turbulence at Re$_{\lambda} = 105$--329 \cite{mhk00}. These data are reanalyzed in the same manner as for our present data. The results are summarized in Fig. \ref{f5}, where quantities are normalized by their values at Re$_{\lambda} = 430$.

The tube radius $R_0$ scales with the Kolmogorov length $\eta$ as $R_0 \propto \eta$ over the entire range of the Reynolds number [Fig. \ref{f5}(a)]. This is the most significant scaling law among those studied here.

The circulation velocity $V_0$ scales with the root-mean-square velocity fluctuation $\langle v^2 \rangle ^{1/2}$ as $V_0 \propto \langle v^2 \rangle ^{1/2}$ at Re$_{\lambda} \gtrsim 400$ [Fig. \ref{f5}(b)]. Since this is not the case at Re$_{\lambda} \lesssim 400$, the scaling is achieved at high Reynolds numbers. Although the velocity fluctuation is a quantity for large scales, vortex tubes could be formed via shear instabilities on the borders of energy-containing eddies \cite{cdc95,vsg95,j93}, where a velocity variation over a small scale such as the tube radius could be comparable to the velocity fluctuation. The circulation velocity also scales with the Kolmogorov velocity $u_K$ as $V_0 \propto u_K$ [Fig. \ref{f5}(c)]. However, at Re$_{\lambda} \gtrsim 400$, this scaling is less significant than the scaling with the velocity fluctuation.

The scaling laws for the radius $R_0$ and the circulation velocity $V_0$ lead to the scaling law for the Reynolds number Re$_0 = R_0 V_0 / \nu$ that characterizes the circulations of vortex tubes:
{\small
\begin{subequations}
\begin{eqnarray}
&{\rm Re}_0& \propto {\rm Re}_{\lambda}^{1/2} \quad
{\rm if} \quad
R_0 \propto \eta 
\quad {\rm and} \quad
V_0 \propto \langle v^2 \rangle ^{1/2}, \\
&{\rm Re}_0& = {\rm constant} \quad
 {\rm if} \quad
R_0 \propto \eta 
\quad {\rm and} \quad
V_0 \propto u_K.
\end{eqnarray}
\end{subequations}
}These relations are from the definitions of the Reynolds number Re$_{\lambda}$, the Kolmogorov length $\eta$, and the Kolmogorov velocity $u_K$ as in Table \ref{t1}. The present result favors the former scaling [Fig. \ref{f5}(d)] rather than the latter [Fig. \ref{f5}(e)] at least for Re$_{\lambda} \gtrsim 400$. With an increase of the Reynolds number Re$_{\lambda}$, vortex tubes have higher Reynolds numbers Re$_0$ and are accordingly more unstable \cite{j93}. They would nevertheless survive long enough to be observable as distinct entities that are responsible for small-scale intermittency. Turbulence is known to be more intermittent at a higher Reynolds number Re$_{\lambda}$ \cite{sa97}.

For general vortex tubes, we do not necessarily expect the scaling laws $V_0 \propto \langle v^2 \rangle ^{1/2}$ and Re$_0 \propto {\rm Re}_{\lambda}^{1/2}$. Weak vortex tubes are not considered here because our velocity profiles were obtained for enhancements of a velocity increment. Actually in direct numerical simulations, the scaling law $V_0 \propto u_K$ was obtained when vortex tubes were identified as local minima of the pressure \cite{kida}. The scaling law $V_0 \propto \langle v^2 \rangle ^{1/2}$ was obtained when vortex tubes were identified as enhancements of the vorticity above a threshold \cite{j93}.

\begin{figure}[!]
\resizebox{8.4cm}{!}{\includegraphics*[3.5cm,8cm][17cm,21cm]{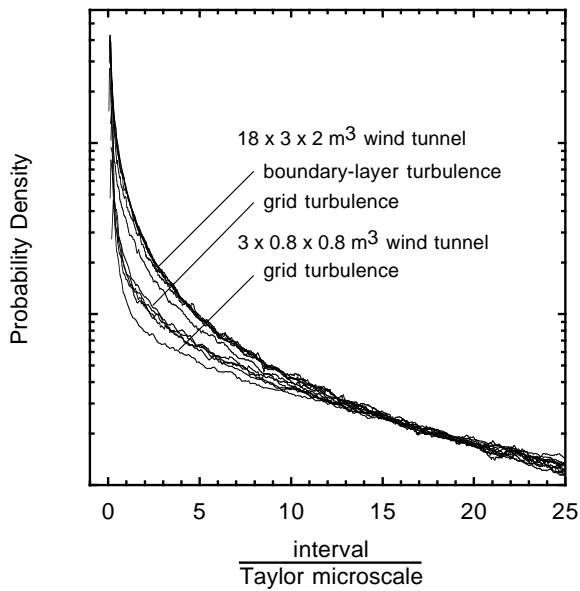}}
\caption{\label{f6} Probability density distribution of the interval $\delta x'$ between vortex tubes in grid turbulence at Re$_{\lambda} = 105$ obtained in a $3 \times 0.8 \times 0.8$ m$^3$ wind tunnel, grid turbulence at Re$_{\lambda} = 165$, 225, 292, and 329 obtained in a $18 \times 3 \times 2$ m$^3$ wind tunnel, and boundary-layer turbulence at Re$_{\lambda} = 295$, 430, 655, 861, 1054, and 1258. We normalize the probability density distribution by the amplitude of its exponential tail. The interval is normalized by the Taylor microscale $\lambda$. The data of boundary-layer turbulence are from our present experiment. The data of grid turbulence are from our previous experiment \cite{mhk00}. Among the data of boundary-layer turbulence, those at Re$_{\lambda} = 295$ yield the lowest probability density.}
\end{figure}

The probability density $P_0$ at the interval $\delta x' = \lambda$ appears to increase as the Reynolds number increases up to Re$_{\lambda} \simeq 400$ [Fig. \ref{f5}(f)]. Above the Reynolds number Re$_{\lambda} \simeq 400$, the probability density is constant. The clustering of vortex tubes appears to become significant and then attain an asymptotic state. There is a discontinuity between grid turbulence at Re$_{\lambda} \lesssim 300$ and boundary-layer turbulence at Re$_{\lambda} \gtrsim 300$. The spatial distribution of vortex tubes is affected by large-scale structures of turbulence, i.e., energy-containing eddies. For large intervals $\delta x' \gg \lambda$, the probability density distributions collapse to single curves according to the experimental configurations as shown in Fig. \ref{f6}.

\section{Concluding Remarks}
\label{s7}

The streamwise ($u$) and spanwise ($v$) velocities were measured simultaneously in rough-wall boundary layers with microscale Reynolds numbers Re$_{\lambda} = 295$--1258. We have used the velocity data to study vortex tubes, the most elementary structures of turbulence.

We have extracted the typical $v$ profile for vortex tubes (Fig.~\ref{f2}). The profile is close to the velocity profile of a Burgers vortex. The radius $R_0$ is several times the Kolmogorov length $\eta$. The maximum circulation velocity $V_0$ is about a half of the root-mean-square velocity fluctuation $\langle v^2 \rangle ^{1/2}$. We have also studied the probability density distribution of the interval between vortex tubes (Fig.~\ref{f4}). The probability density is enhanced below about the integral length and more significantly below about the Taylor microscale, reflecting clustering of vortex tubes.

For the first time at high Reynolds numbers, we have obtained the scaling laws $R_0 \propto \eta$, $V_0 \propto \langle v^2 \rangle ^{1/2}$, and Re$_0 = V_0 R_0 / \nu \propto {\rm Re}_{\lambda}^{1/2}$. The small-scale spatial distribution of vortex tubes is the same (Fig. \ref{f5}). Since these properties do not necessarily exist at Re$_{\lambda} \lesssim 400$, they are achieved asymptotically at Re$_{\lambda} \gtrsim 400$. They are expected to be universal among vortex tubes in turbulence at high Reynolds numbers. To confirm this expectation, experiments at the higher Reynolds numbers are desirable. Those at similar Reynolds numbers but under different experimental configurations are also desirable.

The vortex tubes have been identified using enhancements of a velocity increment above a threshold. Thus our results are biased against weak tubes. The development of a method to identify vortex tubes with various strengths is desirable \cite{mt02}. We nevertheless believe that our results are useful because strong vortex tubes play an important role in small-scale intermittency. Their role in energy dissipation is also expected to be important.

\begin{acknowledgments}
This research has been supported in part by the Japanese Ministry of Education, Science, and Culture under Grant No. (B2) 14340138. The authors are grateful to M. Takaoka for interesting discussions and to E. Kimura for technical support.
\end{acknowledgments}



\begin{thebibliography}{999}

\bibitem{f95} U. Frisch, {\it Turbulence, the Legacy of A.N. Kolmogorov} (Cambridge Univ. Press, Cambridge, 1995), Chap. 8.

\bibitem{sa97} K. R. Sreenivasan and R. A. Antonia, Annu. Rev. Fluid Mech. {\bf 29,} 435 (1997).

\bibitem{s99} K. R. Sreenivasan, Rev. Mod. Phys. {\bf 71,} S383 (1999).


\bibitem{kida} H. Miura and S. Kida, J. Phys. Soc. Jpn. {\bf 66,} 1331 (1997); S. Kida and H. Miura, {\it ibid.} {\bf 69,} 3466 (2000); T. Makihara, S. Kida, and H. Miura, {\it ibid.} {\bf 71,} 1622 (2002). The notion that vortex tubes are the most elementary structures of turbulence was put forward by these authors.

\bibitem{d91} S. Douady, Y. Couder, and M. E. Brachet, Phys. Rev. Lett. {\bf 67,} 983 (1991).

\bibitem{cdc95} O. Cadot, S. Douady, and Y. Couder, Phys. Fluids {\bf 7,} 630 (1995).

\bibitem{vsg95} E. Villermaux, B. Sixou, and Y. Gagne, Phys. Fluids {\bf 7,} 2008 (1995).

\bibitem{lvmb00} A. La Porta, G. A. Voth, F. Moisy, and E. Bodenschatz, Phys. Fluids {\bf 12,} 1485 (2000).

\bibitem{vm91} A. Vincent and M. Meneguzzi, J. Fluid Mech. {\bf 225,} 1 (1991); {\bf 258, } 245 (1994).

\bibitem{j93} J. Jim\'enez, A. A. Wray, P. G. Saffman, and R. S. Rogallo, J. Fluid Mech. {\bf 255,} 65 (1993); J. Jim\'enez and A. A. Wray, {\it ibid.} {\bf 373,} 255 (1998).

\bibitem{tmi99} M. Tanahashi, T. Miyauchi, and J. Ikeda, in {\it IUTAM Symposium on Simulation and Identification of Organized Structures in Flows}, edited by J. N. S$\o$rensen, E. J. Hopfinger, and N. Aubry (Kluwer, Dordrecht, 1999), p. 131.

\bibitem{k03} Exceptionally high Reynolds numbers up to Re$_{\lambda} = 429$ were achieved in the direct numerical simulation of Y. Kaneda, T. Ishihara, M. Yokokama, K. Itakura, and A. Uno, Phys. Fluids {\bf 15,} L21 (2003). The higher Reynolds numbers were achieved by decreasing the maximum wave number from the standard value $2/ \eta$ to a nonstandard value $1/ \eta$. However, such lowered-resolution simulations are not suited to studying vortex tubes.

\bibitem{note0} Vortex tubes with radii that are comparable to the Kolmogorov length $\eta$ do not produce sufficiently low pressure to be observable in bubble visualization, which is thereby biased toward vortex tubes with the larger radii \cite{lvmb00}.  

\bibitem{b96} F. Belin, J. Maurer, P. Tabeling, and H. Willaime, J. Phys. (Paris) II {\bf 6,} 573 (1996). These authors studied velocity data obtained with a single-wire probe in low-temperature helium at Re$_{\lambda} = 151$--5040, and claimed that characteristics of vortex tubes significantly change across Re$_{\lambda} \simeq 700$. Although their results at Re$_{\lambda} \lesssim 700$ are consistent with ours, no significant change of the tube characteristics across Re$_{\lambda} \simeq 700$ has been found in our experiment.


\bibitem{n97} A. Noullez, G. Wallace, W. Lempert, R. B. Miles, and U. Frisch, J. Fluid Mech. {\bf 339,} 287 (1997).

\bibitem{cg99} R. Camussi and G. Guj, Phys. Fluids {\bf 11,} 423 (1999).

\bibitem{mtk99} H. Mouri, M. Takaoka, and H. Kubotani, Phys. Lett. A {\bf 261,} 82 (1999).

\bibitem{mhk00} H. Mouri, A. Hori, and Y. Kawashima, Phys. Lett. A {\bf 276,} 115 (2000); Phys. Rev. E {\bf 67,} 016305 (2003).

\bibitem{mt02} H. Mouri and M. Takaoka, Phys. Rev. E {\bf 65,} 027302 (2002). 

\bibitem{note1} The 99\% thickness and displacement thickness are usually defined not with the maximum value $\hat{U}$ of the mean streamwise velocity but with the incoming-wind velocity $U_i$. However, since our wind tunnel was not capable of adjusting its ceiling to set the streamwise pressure gradient to be zero, the mean streamwise velocities at large heights were greater than the incoming-wind velocity. We thereby used the maximum value of the mean streamwise velocity obtained at $\hat{z} = 0.90$ m.

\bibitem{mthk03} H. Mouri, M. Takaoka, A. Hori, and Y. Kawashima, Phys. Rev. E {\bf 68,} 036311 (2003).

\bibitem{b95} H. H. Bruun, {\it Hot-Wire Anemometry, Principles and Signal Analysis} (Oxford Univ. Press, Oxford, 1995), Chaps. 2 and 5.

\bibitem{note2} The two wires individually respond to all the $u$, $v$, and $w$ components. Since the measured $u$ component corresponds to the sum of the responses of the two wires, it is contaminated with the $w$ component. The contamination is measured as an increase of the streamwise velocity. Since the measured $v$ component corresponds to the difference of the responses, it is free from the $w$ component.

\bibitem{note3} Despite the difference in the Reynolds number Re$_{\lambda}$, the probability density distributions of the velocity increment in Fig. \ref{f3} are nearly the same. This is because the energy-containing scale, which is comparable to the integral length $L$, and the scale $\delta x$ for the velocity increment are nearly the same. The scale range for the energy cascade is nearly the same.

\bibitem{s02} K. Sassa, in {\it Generation-Sustenance Mechanism and Statistical Law of Turbulence,} edited by S. Kida and S. Goto (Research Institute for Mathematical Sciences, Kyoto, 2002), p. 154 (in Japanese).

\bibitem{note5} The observed positive excursion is partially attributable to fluctuation of the velocity $U+u$ at which a vortex tube passes the probe. Under Taylor's frozen-eddy hypothesis, the velocity increment over the given small scale is more enhanced for a faster-moving tube, which is more likely to be incorporated in our conditional averaging \cite{mhk00}.

\bibitem{t92} A. Tsinober, E. Kit, and T. Dracos, J. Fluid Mech. {\bf 242,} 169 (1992); M. Kholmyansky, A. Tsinober, and S. Yorish, Phys. Fluids {\bf 13,} 311 (2001).

\bibitem{note4} While experimental curves represent mere loci of discrete data points in Figs. \ref{f1}--\ref{f3}, smoothing has been applied to the tails of the probability density distributions in Figs. \ref{f4} and \ref{f6}.

\bibitem{f68} W. Feller, {\it An Introduction to Probability Theory and Its Applications,} 3rd ed. (Wiley, New York, 1968), Vol. 1, Chap. 17.

\end{thebibliography}
\end{document}